\documentstyle[11pt,epsf]{article}
\newlength{\bredde}
\def\slash#1{\settowidth{\bredde}{$#1$}\ifmmode\,\raisebox{.15ex}{/}
\hspace*{-\bredde} #1\else$\,\raisebox{.15ex}{/}\hspace*{-\bredde} #1$\fi}
\newcommand{\beq}{\begin{equation}}
\newcommand{\eeq}{\end{equation}}
\newcommand{\noi}{\vspace{12pt}\noindent}
\newcommand{\lG}{\raise.3ex\hbox{$\stackrel{\leftarrow}{G}$}}
\newcommand{\lU}{\raise.3ex\hbox{$\stackrel{\leftarrow}{U}$}}
\newcommand{\lP}{\raise.3ex\hbox{$\stackrel{\leftarrow}{{\cal P}}$}}
\newcommand{\leta}{\raise.3ex\hbox{$\stackrel{\leftarrow}{\eta}$}}
\newcommand{\lOmega}{\raise.3ex\hbox{$\stackrel{\leftarrow}{\Omega}$}}
\newcommand{\ldr}{\raise.3ex\hbox{$\stackrel{\leftarrow}{\delta^r}$}}

\def\beqn{\begin{eqnarray}}
\def\eeqn{\end{eqnarray}}

\def\gtwid{\raise.3ex\hbox{$>$\kern-.75em\lower1ex\hbox{$\sim$}}}
\def\ltwid{\raise.3ex\hbox{$<$\kern-.75em\lower1ex\hbox{$\sim$}}}

\begin{document}
\topmargin -1.4cm
\oddsidemargin -0.8cm
\evensidemargin -0.8cm
\title{\Large{{\bf Topology and the Dirac Operator Spectrum \\ in
Finite-Volume Gauge Theories}}}

\vspace{1.5cm}

\author{~\\~\\
{\sc P.H. Damgaard}\\
The Niels Bohr Institute\\ Blegdamsvej 17\\ DK-2100 Copenhagen\\
Denmark}
 
\maketitle
\vfill
\begin{abstract} The interplay between between gauge-field winding numbers,
$\theta$-vacua, and the Dirac operator spectrum in finite-volume gauge
theories is reconsidered. To assess the weight of each topological
sector, we compare the mass-dependent chiral condensate in 
gauge field sectors of fixed topological index with the answer obtained
by summing over the topological charge. Also the microscopic Dirac operator 
spectrum in the full finite-volume 
Yang-Mills theory is obtained in this way, by summing over all topological 
sectors with the appropriate weight. 
\end{abstract}
\vfill
\begin{flushleft}
NBI-HE-99-07 \\
hep-th/9903096
\end{flushleft}
\thispagestyle{empty}
\newpage

\setcounter{page}{1}
\section{Introduction}

Although the chiral SU$_L(N_f)\times$SU$_R(N_f)$ symmetry of QCD is not 
spontaneously broken in a finite volume, it has long been known that
many low-energy aspects of finite-volume QCD nevertheless can be derived 
from an effective chiral Lagrangian \cite{GL}. This is because the
dominating degrees of freedom can still be the ``would-be'' 
$N_f^2-1$ pseudo-Goldstone bosons (which are simply called pions in what 
follows) of the presumed chiral symmetry breaking pattern of
SU$_L(N_f)\times$SU$_R(N_f) \to$ SU($N_f$) in the infinite volume limit.
Clearly at least two scales are of relevance here: if the linear 
extent $L$ of the four-volume $V$ is much larger than the Compton wavelength
$1/m_{\pi}$ of the pions, the effect of the finite volume is mild, 
and can be treated
as a small correction to the low-energy results of the theory in infinite
volume. What is at first sight much more surprising is the fact that
even in the opposite limit, when the pion Compton wavelenght much {\em
exceeds} the size of the box, the effective partition function still becomes
of a very simple chiral Lagrangian type. The procedure to derive this
effective theory is to treat the zero modes of the pion field as new
collective coordinates in the low-energy expansion. When all the dust has
settled, the resulting leading-order Lagrangian coincides with
what one would have obtained by the naive procedure of simply discarding
all derivative terms in the infinite-volume chiral Lagrangian \cite{GL}.
Independently of whether chiral symmetry is spontanenously broken or not,
the pseudoscalar states we here generically denote by pions are 
guaranteed, by exact mass inequalities, to be the lightest flavor
non-singlet mesons in QCD \cite{W}.

\noi
Naively one might think that an effective Lagrangian technique as described
above can provide information only about the chiral order parameter
$\Sigma \equiv \langle\bar{\psi}\psi\rangle$ itself, or just soft pion 
dynamics. However, as shown by Leutwyler and Smilga \cite{LS}, low-energy 
constraints on finite-volume gauge theories can, remarkably, be used to 
provide very non-trivial information about quark and gluon degrees of freedom
as well. Specifically, the large-volume identification of the original
QCD partition function and the effective Lagrangian restricted to the
region 
\beq
\frac{1}{\Lambda_{QCD}} ~\ll~ L ~\ll~ \frac{1}{m_{\pi}} ~, 
\label{meso}
\eeq
carries a lot of exact
information about the spectral correlators of the smallest eigenvalues of
the Dirac operator. While this holds for the full partition functions 
involved, the
scoop of ref. \cite{LS} was to note that the analysis simplifies considerably
if one focuses on gauge field sectors of fixed topological charge $\nu$.
The effective partition function then becomes analytically solvable. 
It is thus possible to do
a direct analytical comparison between, on one side, the full QCD
partition function, and, on the other side, the effective partition function.
One of the most striking outcomes of this analysis was the derivation
of exact spectral sum rules for the Dirac operator in QCD and QCD-like
gauge theories \cite{LS,SmV}.    

\noi
Surprisingly, the analysis of the spectral properties
of the Dirac operator in finite-volume gauge field theories at fixed
topological charge \cite{LS} has turned out to have an alternative description
in terms of Random Matrix Theory \cite{SV,V}. The root of this equivalence
lies in the universality of the Random Matrix Theory predictions: there
is a huge class of Random Matrix Theories that, in the appropriate
scaling limit, yield {\em exactly} the same spectral correlation functions
\cite{ADMN,DN,SeV,S}. Remarkably, so-called chiral versions of the
three classical Random Matrix Theory ensembles have been shown by
Verbaarschot to provide precisely the needed three universality classes of 
chiral symmetry breaking in gauge theories with different gauge groups
and different color representations of the fermions \cite{Jac}. While
there are obvious resemblances here to the classification of critical behavior
in terms of universality classes in statistical mechanics, the analogy
should not be pushed too far. There is no renormalization group at work
here, and the way QCD fits into one of these three universality classes
is by means of algebra (or, at a deeper level, symmetry) rather than dynamics.
Indeed, within the last year it has become well understood how all the 
universal predictions derived from Random Matrix Theory can be rephrased in 
terms of the effective QCD partition function, suitably extended with 
additional quark species \cite{D0}. Even universality of the results can 
easily be understood directly from the universality of the Random Matrix 
Theory partition functions \cite{D1} and the equivalence between
the effective chiral Lagrangian and the specific Random Matrix Theory with
gaussian distribution, which was established in \cite{SV}. 
By means of partially 
quenched chiral Lagrangians, it is now also possible to derive directly from 
the effective QCD partition function the microscopic spectral density of 
the Dirac operator, without any bypass through Random Matrix Theory at all 
\cite{OTV}. All these more recent results permit us to say much more about 
the microscopic Dirac operator spectrum in finite-volume gauge theories. It 
is therefore appropriate to reconsider the analysis of ref. \cite{LS} 
in this new light. 

\noi
It is thus partly by historical coincidence that almost all
analytical predictions about the microscopic Dirac operator spectrum and
the mass-dependent chiral condensate have been derived in gauge sectors of 
fixed winding number $\nu$. From the point of view of the ``real'' theory,
where one sums over all topological sectors, this is analogous to knowing
only the Fourier modes of a periodic function. While it may be interesting
to study just these Fourier modes (and in the case of QCD much can indeed
be learned from them), we would still rather like to know the full function.
For QCD this means that we should know the predictions for the full theory
at any given value of the vacuum angle $\theta$, or, at least, since there
are very stringent experimental bounds on the magnitude of this CP-violating
$\theta$-angle, for the specific value of $\theta=0$.

\noi
The problem of summing over all winding numbers becomes particularly 
important when we consider tests of
the analytical predictions for QCD sectors of fixed topological charge
$\nu$ by means of lattice gauge theory. Because lattice gauge configurations
are not smooth on the scale of the lattice spacing, the very definition of 
gauge field topology is from the
beginning an ambiguous concept. With so-called staggered fermions there is 
no help from index theorems (which otherwise can serve as an alternative 
definition of
topological sectors) because there are no guaranteed exact zero modes away 
from the continuum limit \cite{SVink}. In fact, lattice gauge theory data for
QCD with staggered fermions {\em and no restrictions on lattice field
topology} show remarkably good agreement with the analytical predictions
for just the sector of zero winding number, $\nu=0$. This phenomenon
was first observed in comparisons of lattice data with analytical formulas
for the mass-dependent chiral condensate \cite{Jac1}\footnote{Based on
lattice data with dynamical quark masses so large on the relevant scale
that the simulations effectively were quenched.}, 
and later precisely
the same effect was seen in lattice simulations of the microscopic Dirac
operator spectrum with staggered fermions \cite{BBetal,DHK}. One can
argue that just because staggered fermions have no
exact zero modes, this explains the puzzle. Are present-day 
lattice gauge theory simulations with staggered fermions thus only 
probing the sector of zero topological charge?\footnote{In this connection, 
see the recent careful studies by the SCRI and Columbia groups \cite{Chen}.} 
Despite the problem
mentioned above, lattice gauge field configurations 
generated by conventional Monte Carlo algorithms cannot all be classified as
topologically trivial. They contain a distribution of configurations that
smoothly fall into the precise topological sectors of the continuum theory
as the lattice cut-off is decreased. Of course, in a given Monte Carlo
sampling the sector corresponding to $\nu=0$ will be dominant. To understand
the significance of just these configurations
we need to know not only the analytical predictions for given sectors of
{\em fixed} topological index, but also the weight with which these different
topological sectors are contributing to full QCD. Very recently it has
become feasible to address this in lattice gauge theory
\cite{L,Urs}. But in the finite-volume region (\ref{meso}), 
these questions can also be answered
analytically. For QCD-like theories the relevant partition functions at 
fixed topological charge $\nu$ are known in closed analytical form
\cite{JSV}. It is simple to show that these fixed-$\nu$ partition functions 
precisely give the non-trivial weights with which each fixed-$\nu$ observable
contributes to the full answer. Using this fact, we assess in section 2
the importance of the $\nu=0$ sector in bulding up the mass-dependent
chiral condensate of the full theory. Section 3 is devoted to the 
microscopic spectral density of the Dirac operator, and we carry out the 
analogous procedure of summing over topological charges. Interestingly,
one of the simple conclusions is that in the {\em massless} case the 
microscopic
spectral density of the Dirac operator in the full theory simply equals
that of the $\nu=0$ sector. This holds for all $N_f \geq 2$. The case
$N_f=1$ is -- as expected -- special, and there are then additional
zero-mode contributions to the full microscopic spectral density from just the
two sectors of $\nu=+1$ and $\nu=-1$. In the massless limit just these two
zero-mode contributions to the spectral density provide the required
non-vanishing condensate for $N_f=1$, even in this finite-volume situation. 
By numerically performing the sum over 
topological charges, we show how the full microscopic spectral density changes
away from the $\nu=0$ prediction as the mass is increased for different 
values of $N_f$. We point out a difficulty with taking the quenched limit,
and briefly discuss the more exotic chiral
symmetry breaking cases, which correspond to different symmetries of the
Dirac operator. All of the predictions described here can be 
tested in lattice gauge theory simulations. In fact, one of our main points
is precisely that one can simultaneously test both the detailed analytical 
predictions in fixed-$\nu$ sectors, {\em and} the predicted weight with 
which these sectors build up the full answer (which, in principle, should be
much easier to measure). Section 5 contains our conclusions.

\section{Summing over Winding Numbers} 

\noi
We begin with a few basic definitions. Consider QCD in Euclidean space on
a finite 4-dimensional torus. Its four-volume can be taken to be just
$L^4$, or $L^3/T$. The former case leads in the $L\to\infty$ limit to ordinary 
zero-temperature QCD, while in the latter case, if we take $T$ fixed,
the $L \to \infty$ limit gives Euclidean QCD at (small\footnote{The 
temperature must be bounded by $m_{\pi} \ll T \ll \Lambda_{QCD}$, 
in complete analogy with the condition (\ref{meso}).}) temperature $T$. 
For our purposes there is no need to distinguish between the two cases, as
the infinite (three-)volume limit of the chiral condensate just must
be chosen appropriately. The four-volume is in both cases denoted by $V$.
As discussed at length in $e.g.$ refs. \cite{LS,'t}, gauge fields on such
a four-torus can be described by a gauge potential $A_{\mu}(x)$
on the full Euclidean space, plus a transition function $\Omega(x)$,
which describes the transition from one periodic cell to another (possibly
non-trivial because $A_{\mu}(x)$ need only be periodic modulo gauge 
transformations). On the lattice one often prefers to work with strictly 
periodic gauge link variables (and hence trivial transition function), but by 
a gauge transformation these links can be made equivalent to a set of link 
variables with a non-trivial transition function. In the continuum such a 
gauge transformation will, however, necessarily be singular. Non-trivial 
winding numbers arise precisely from non-trivial transition functions 
\cite{'t}.

\noi
For an SU($N_c$) gauge theory with $N_f$ fermions in the fundamental
representation, antiperiodic boundary conditions on the four-torus for these
fermions the topological charge
\beq
\nu ~=~ \frac{1}{32\pi^2} \int\! d^4x {\mbox{\rm Tr}}
~[F_{\mu\nu}F_{\rho\sigma}]\epsilon_{\mu\nu\rho\sigma}
\eeq
runs over integer values. With a $\theta$-term in the action,
the partition function thus reads
\beq
{\cal Z}(\theta) ~=~ \sum_{\nu=-\infty}^{\infty}
e^{i\nu\theta}\int\![dA]_{\nu}\prod_f \det[i\slash{D} - m_f] e^{S_{YM}[A]} ~,
\label{ZQCD}
\eeq
after integrating out the fermions. Alternatively,
\beq
{\cal Z}(\theta) ~=~  \sum_{\nu=-\infty}^{\infty}e^{i\nu\theta} {\cal Z}_{\nu} ~,
\eeq
where ${\cal Z}_{\nu}$ is like a Fourier transform of the full
partition function:
\beq
{\cal Z}_{\nu} ~=~ \frac{1}{2\pi}\int_0^{2\pi}\!d\theta e^{-i\nu\theta}
{\cal Z}(\theta) ~.\label{Znudef}
\eeq

\noi
Consider first the case $N_f\geq 2$, and assume that chiral symmetry in
the infinite-volume limit breaks spontaneously according to
SU$_L(N_f)\times$SU$_R(N_f) \to$ SU($N_f$). As shown in ref \cite{GL}, 
the effective partition function then takes on a very simple form in the 
finite-volume region (\ref{meso}):
\beq
{\cal Z}(\theta) ~=~ \int_{SU(N_{f})}\! dU
\exp\left[V\Sigma {\mbox{\rm Re}}\,[e^{i\theta/N_{f}}{\mbox{\rm Tr}}\,
{\cal M}U^{\dagger}]\right] ~, \label{Zeff}
\eeq
where ${\cal M}$ is the quark mass matrix\footnote{{}From now on we
always take this mass matrix to be real and diagonal: ${\cal M}$ = 
diag($m_1,\ldots,m_{N_{f}}$).}, and $\Sigma$ is the infinite-volume
chiral condensate (for the massless theory with $\theta=0$). 
One of the main lessons
to learn from the analysis of ref. \cite{LS} is that these
finite-volume effective partition functions for $N_f\geq 2$ in surprisingly
many ways resemble the finite-volume partition function with $N_f=1$.
In fact, although the case $N_f=1$ is radically different in that the
axial U(1) symmetry is broken by the anomaly rather than spontaneously,
the effective partition function turns out be given by what
one naively could have extrapolated from (\ref{Zeff}), namely
the simple exponential \cite{LS} 
\beq
{\cal Z}(\theta) ~=~ \exp\left[V\Sigma{\mbox{\rm Re}}
[e^{i\theta}m]\right] ~.
\eeq
In all cases the effective partition functions depend
only on the combination $\mu_i \equiv \Sigma V m_i$. Keeping $\mu_i$ fixed
as the four-volume $V$ is taken to infinity thus entails taking the chiral
limit in a correlated manner. Moreover, the bound (\ref{meso}) must be
satisfied throughout. In statistical mechanics such a procedure is known
as finite-size scaling, and the $\mu_i$ are finite-size scaling variables.

\noi
The Fourier coefficients ${\cal Z}_{\nu}$ have been computed analytically
for all $N_f$ \cite{LS,JSV}. For $N_f\geq 2$ they can conveniently be written
\beq
{\cal Z}_{\nu}(\{\mu_i\}) ~=~ \frac{\det {\cal A}(\{\mu_i\})}{\Delta(\mu^2)} ~,
\label{Zeffnu}
\eeq
where the $N_f\times N_f$ matrix ${\cal A}(\{\mu_i\})$ is ($I_n(x)$ is the
modified Bessel function of order $n$) 
\beq
{\cal A}(\{\mu_i\}) ~=~ \mu_i^{j-1}I_{\nu+j-1}(\mu_i) ~,
\eeq
and
\beq
\Delta(\mu^2) ~=~ \prod_{i<j}(\mu_i^2-\mu_j^2)
\eeq
is the Vandermonde determinant. With an obvious interpretation of the
formula (\ref{Zeffnu}), it even includes the case $N_f=1$, for which \cite{LS}
\beq
{\cal Z}_{\nu}(\mu) ~=~ I_{\nu}(\mu) ~.
\label{ZNf1}
\eeq
One notices that the general expression (\ref{Zeffnu}) is symmetric in $\nu$,
\beq
{\cal Z}_{\nu}(\{\mu_i\}) ~=~  {\cal Z}_{-\nu}(\{\mu_i\}) ~,
\label{Zsym}
\eeq
as follows also directly from the original definition, eqs. (\ref{Znudef}) and
(\ref{Zeff}).
 
\subsection{\sc The chiral condensate for $N_f=1$}

\noi
Having already observed that in the limit (\ref{meso}) the relevant 
scaling variables are $\mu_i = \Sigma V m_i$, 
we from now on try to eliminate as many factors of
$V$ and $\Sigma$ as possible. We hence measure all mass-dependent chiral
condensates $\Sigma(\theta;\{\mu_i\})$ in units of $\Sigma$, and effectively
work with a unit four-volume. The mass-dependent chiral condensate for
$N_f=1$ is then simply defined by
\beq
\Sigma(\theta;\mu) ~=~ \frac{\partial}{\partial\mu} \ln{\cal Z}(\theta;\mu) ~.
\eeq
In this $N_f=1$ case the result is trivial and in fact $\mu$-independent: 
$\Sigma(\theta;\mu) = \cos(\theta)$, 
as follows from eq. (\ref{ZNf1}).\footnote{The $\theta$-dependence
of the condensate is not interesting; for $\theta \neq 0$ it simply
shifts from $\langle\bar{\psi}\psi\rangle$ to $\langle\bar{\psi}\gamma^5
\psi\rangle$ \cite{LS}.} In any case, we shall mainly be concerned with the 
case $\theta=0$, for which, in our units, $\Sigma(0;\mu)=1$.

\noi
Let us now compare with what we would find if we were to restrict ourselves
to gauge field averages over fixed topological charge only. As already 
observed by Leutwyler and Smilga {\cite{LS}, this restriction is quite 
drastic. By definition,
\beq
\Sigma_{\nu}(\mu) ~=~ \frac{\partial}{\partial\mu} \ln{\cal Z}_{\nu}(\mu) ~,
\label{sigmanu1def}
\eeq
which gives
\beq
\Sigma_{\nu}(\mu) ~=~ \frac{1}{I_{\nu}(\mu)}\left[I_{\nu+1}(\mu) +
\frac{\nu}{\mu}I_{\nu}(\mu)\right] ~. 
\label{sigmanu1}
\eeq
For $\nu$ positive, the last term can easily be traced to the $|\nu|$ zero 
modes of the Dirac operator $\slash{D}$: In terms of the non-zero Dirac 
eigenvalues $\lambda_n$ the original partition function (\ref{ZQCD}) reads
\beq
{\cal Z}(\theta) ~=~ \sum_{\nu=-\infty}^{\infty}\prod_f(m_f)^{|\nu|}
e^{i\nu\theta}\int\![dA]_{\nu}\prod_f \prod_n[\lambda_n^2 + m_f^2] 
e^{S_{YM}[A]} ~,
\label{ZQCDeig}
\eeq
and the last term in (\ref{sigmanu1}) precisely matches the term due to
the prefactor $m^{|\nu|}$. However, when $\nu$ is negative this interpretation
is not correct: in that case the sign of the second term in (\ref{sigmanu1})
is just the opposite of what is needed. There is no mystery here, but
only a slight complication which we shall return to in section 3.1. Indeed, 
from the very definition (\ref{sigmanu1def}) and the property (\ref{Zsym}),
the mass-dependent chiral condensate (\ref{sigmanu1}) is seen to be
symmetric in $\nu$: $\Sigma_{\nu}(\mu)=\Sigma_{-\nu}(\mu)$. Therefore
the obvious $\sim 1/\mu$ singularity in eq. (\ref{sigmanu1}) for $\nu > 0$
is actually matched by a singularity of exactly the same strength
for $\nu < 0$.

\noi
We start by considering the case $\theta=0$. 
In fig. 1 we show $\Sigma_{\nu}(\mu)$
for a few small values of $\nu$, and compare each of them with the full
answer $\Sigma(\mu)$. Contrary to the full answer which in this 
finite-volume limit
is just the trivial constant $\Sigma=1$, the chiral 
condensates in individual fixed-$\nu$ sectors have non-trivial mass
dependences. In this case we certainly cannot conclude that the $\nu=0$
sector adequately reproduces the full answer. But fixed-$\nu$ chiral
condensates also show another disturbing feature, which we already mentioned
above: for $\nu\neq 0$ they {\em diverge}
in the limit $\mu\to 0$. Actually, the second term in eq. (\ref{sigmanu1})
by itself does not contribute to the full sum over topological sectors.
To see this,
we first derive a simple formula that tells us how to relate the
full chiral condensate $\Sigma(\theta;\mu)$ to the chiral condensates
$\Sigma_{\nu}(\mu)$ in sectors of fixed topological index:
\begin{eqnarray}
\Sigma(\theta;\mu) &=& \frac{\partial}{\partial\mu}\ln {\cal Z}(\theta;\mu)\cr
&=& \frac{\partial}{\partial\mu}\ln\left(\sum_{\nu=-\infty}^{\infty} 
{\cal Z}_{\nu}(\mu)e^{i\nu\theta}\right)\cr
&=& \frac{1}{{\cal Z}(\theta;\mu)}\sum_{\nu=-\infty}^{\infty}e^{i\nu\theta} 
\frac{\partial}{\partial\mu}{\cal Z}_{\nu}(\mu)\cr
&=& \frac{1}{{\cal Z}(\theta;\mu)}\sum_{\nu=-\infty}^{\infty}e^{i\nu\theta} 
{\cal Z}_{\nu}(\mu)\Sigma_{\nu}(\mu)\cr
&=& \langle\langle\Sigma_{\nu}(\mu)\rangle\rangle ~.
\label{sigmasum}
\end{eqnarray}
We have here defined an ``average over topology'' by
\beq
\langle\langle F(\nu)\rangle\rangle ~\equiv~ \frac{1}{{\cal Z}(\theta;\mu)}
\sum_{\nu=-\infty}^{\infty}e^{i\nu\theta}{\cal Z}_{\nu}(\mu)F(\nu) ~.
\label{topave}
\eeq

\centerline{\epsfbox{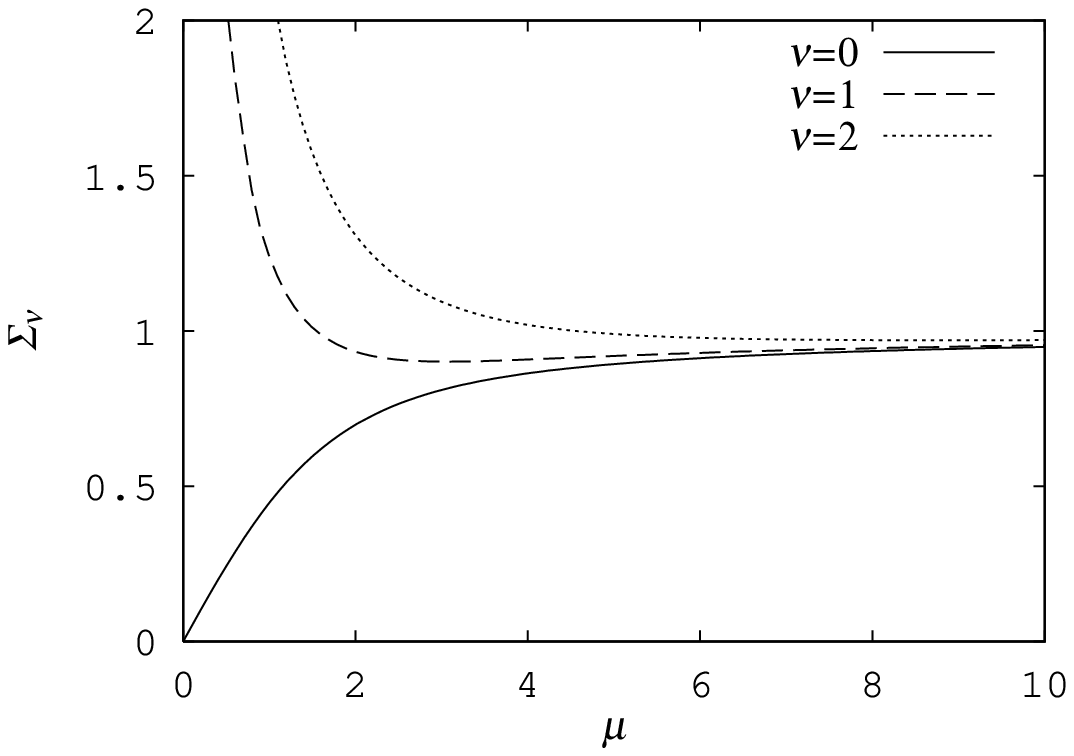}}
{{\small Figure 1: The mass-dependent chiral condensate 
$\Sigma_{\nu}(\mu)$ for $N_f=1$. Shown are the results for a few fixed 
values of $\nu$. Even for $\theta=0$ they are quite different from the
full answer (the constant 1 in that case).}}

\noi
~

\noi
The relation (\ref{sigmasum}) tells us with which weight the chiral 
condensates in fixed topological sectors contribute to the full answer: Each 
$\Sigma_{\nu}(\mu)$ comes with weight factor $e^{i\nu\theta}Z_{\nu}(\mu)$.
In fig. 2 we show how the full answer, the constant $\Sigma = 1$ for
$\theta=0$, is built 
up as we sum over topological charge $\nu$ from a given $-|\nu|_{max}$ to
$+|\nu|_{max}$ in eq. (\ref{sigmasum}). Perhaps surprisingly, 
we find that the $\nu=0$ contribution
alone is not very significant, never exceeding more than around 20\% of
the full answer. In the shown range of $\mu$ we also see that by summing 
from, say, $\nu=-8$ to $\nu=+8$ we essentially obtain the complete $\Sigma$.
The divergences that appear in $\Sigma_{\nu}(\mu)$ for $\nu\neq 0$ are
obviously completely removed when one performs the required average over
topology. In the present case of $\theta=0$ this can easily be 
seen analytically. According to eqs. (\ref{sigmanu1}) and (\ref{sigmasum}), 
and using ${\cal Z}(\theta\!=\!0;\mu)=\exp[\mu]$, we here have
\beqn
\Sigma(\theta=0;\mu) &=& e^{-\mu}\sum_{\nu=-\infty}^{\infty}\left[
I_{\nu+1}(\mu)+\frac{\nu}{\mu}I_{\nu}(\mu)\right]\cr
&=& e^{-\mu}\sum_{\nu=-\infty}^{\infty}I_{\nu+1}(\mu) ~.\label{simple}
\eeqn

\centerline{\epsfbox{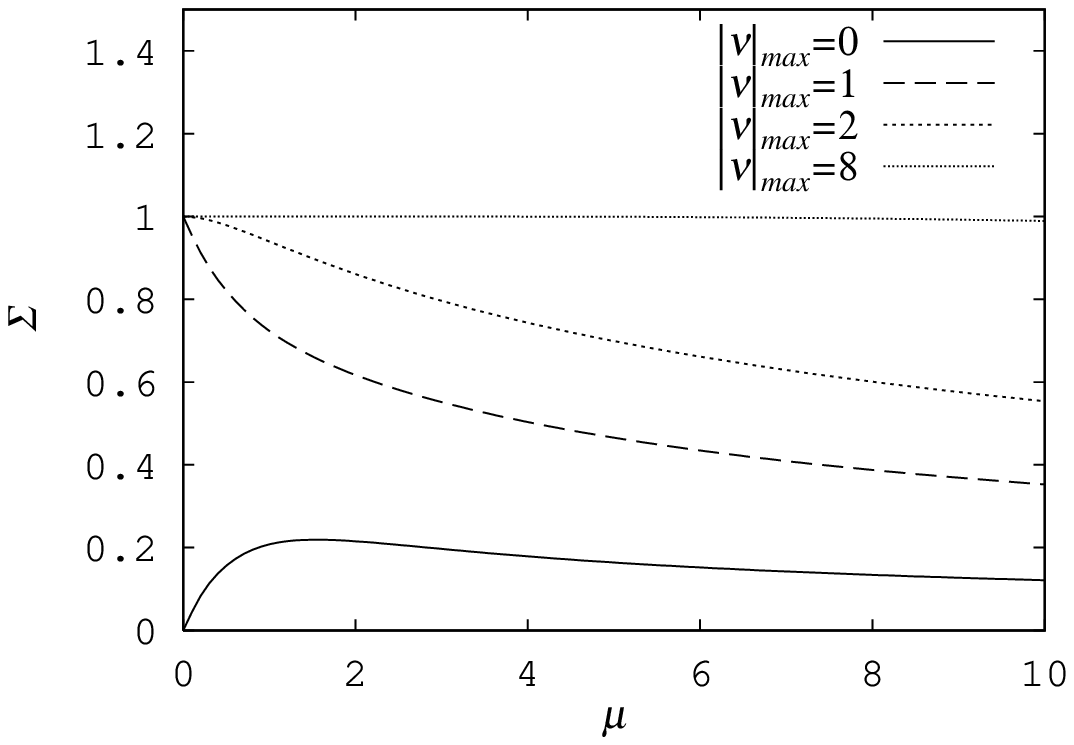}}
{{\small Figure 2: The full mass-dependent chiral condensate 
$\Sigma(\theta\!=\!0,\mu)$ for $N_f=1$. The exact answer equals 1.
Shown are the results of summing over the first few  
values of $\nu$. The $\nu=0$ sector is not particularly significant.}}

\noi
~

\noi
The second term in the first line actually does not contribute at all:
This is obvious for $\nu=0$,
and for $\nu\neq 0$ the contribution from $+\nu$ precisely cancels the
contribution from $-\nu$. Next, substituting the Bessel function identity
\beq
\sum_{\nu=-\infty}^{\infty}e^{i\nu\theta}I_{\nu}(\mu) ~=~ 
e^{\mu\cos(\theta)} ~,\label{Bid}
\eeq
we of course recover the correct result $\Sigma(\theta=0;\mu)=1$ from
eq. (\ref{simple}).

\noi
For a non-zero vacuum angle $\theta$ the $\nu/\mu$-term in
$\Sigma_{\nu}(\mu)$ does not disappear as simply when we sum over different
topological sectors. Again this $N_f=1$ case is very illustrative. Now,
using some simple Bessel function identities, we have 
\beqn
\Sigma(\theta;\mu) &=& e^{-\mu\cos(\theta)}\sum_{\nu=-\infty}^{\infty}
e^{i\nu\theta}\left[I_{\nu+1}(\mu)+\frac{\nu}{\mu}I_{\nu}(\mu)\right]\cr
&=& \frac{1}{2}e^{-\mu\cos(\theta)}\sum_{\nu=-\infty}^{\infty}
e^{i\nu\theta}\left[I_{\nu+1}(\mu) + I_{\nu-1}(\mu)\right]\cr
&=& \cos(\theta) ~.\label{thetanot0}
\eeqn
where in the last line we have agin used eq. (\ref{Bid}).
In this case the second term in (\ref{sigmanu1}) is certainly required to 
obtain the
right answer from the sum (\ref{sigmasum}). More generally, we also learn 
from eq. (\ref{sigmasum}) that
the $\mu\to 0$ singularities in $\Sigma_{\nu}(\mu)$ for 
$\nu\neq 0$ are killed in the sum over topological sectors because these
divergences are weighted by $e^{i\nu\theta}Z_{\nu}(\mu)$. In general,
$Z_{\nu}(\mu) \sim \mu^{N_{f}|\nu|}$ for small $\mu$. This follows both
explicitly from the representation (\ref{Zeff}), or, in the original QCD
language, from the Dirac determinant and the index theorem. 
Thus only for $N_f=1$,
and in that case only for the sectors with $\nu=\pm 1$, can these $1/\mu$
singularities in $\Sigma_{\nu}(\mu)$ give any non-zero contribution to the 
massless chiral condensate. In fact, one easily checks that in this case 
just these two contributions precisely yield the finite value 
$\Sigma(\theta;0)=\cos(\theta)$, as required \cite{LS}. 

\subsection{\sc More flavors: $N_f \geq 2$.}

\noi
It is convenient to first restrict ourselves to the case of equal masses,
$\mu_i = \mu, ~\forall i$, so that the chiral condensates of all $N_f$
species become equal. Then
\beq
\Sigma(\theta;\mu) ~\equiv~ \frac{1}{N_{f}}\frac{\partial}{\partial\mu}
\ln {\cal Z}(\theta;\mu) ~=~ \langle\langle \Sigma_{\nu}(\theta;\mu)
\rangle\rangle ~, \label{sigmanfdef}
\eeq 
with $\Sigma_{\nu}(\theta;\mu)$ defined analogously. Consider first the 
case $N_f=2$. The full partition function is known analytically \cite{LS}:
\beq
{\cal Z}(\theta;\mu) ~=~ \frac{\sqrt{2}}{\mu\sqrt{(1+\cos(\theta))}}
I_{1}(\mu(2(1+\cos(\theta))^{1/2}) ~.\label{Z2}
\eeq

\centerline{\epsfbox{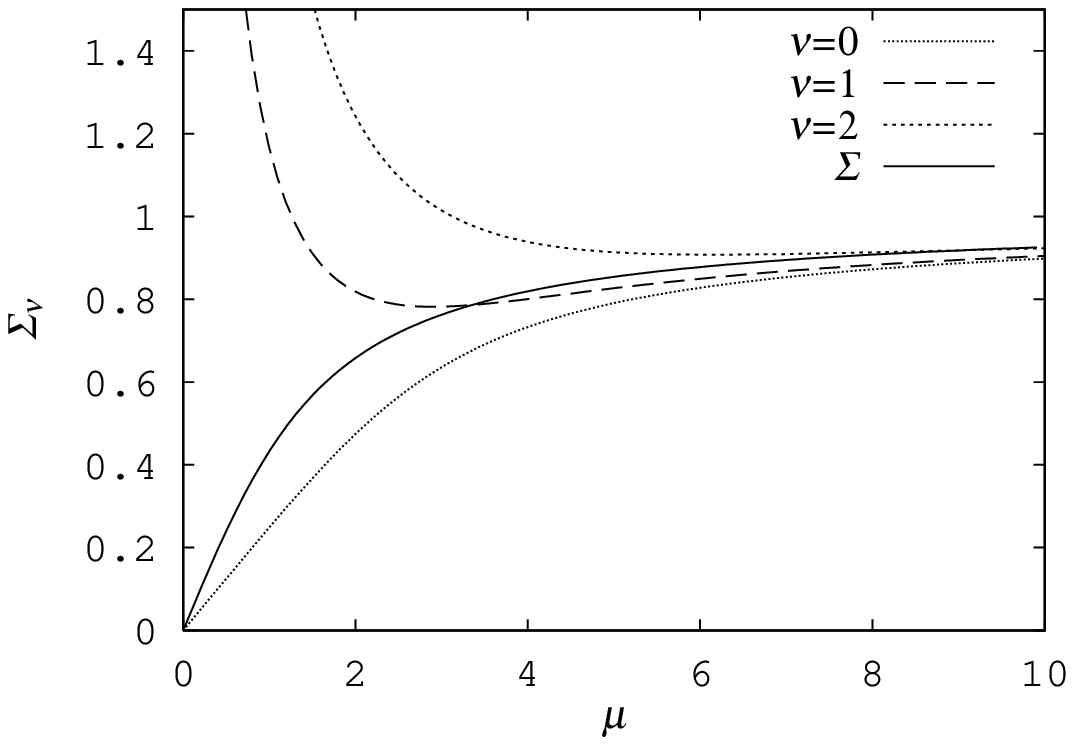}}
{{\small Figure 3: The chiral condensate 
$\Sigma_{\nu}(\mu)$ for $N_f=2$ and equal masses $\mu$. 
There is no qualtitative difference with the picture from the $N_f=1$
theory. For comparison we have also displayed the full chiral condensate
$\Sigma(\mu)$ for $\theta=0$.}}

\noi
~

\noi
Let us again first restrict ourselves to the case $\theta=0$. The full 
mass-dependent chiral condensate then becomes
\beq
\Sigma(\theta=0;\mu) ~=~ \frac{I_0(2\mu)+I_2(2\mu)}{2I_1(2\mu)}
-\frac{1}{2\mu} ~,\label{sigma2}
\eeq
which, despite appearances, is not singular at $\mu=0$. The partition
function (\ref{Zeffnu}) reads
\beq
Z_{\nu}(\mu) ~=~ I_{\nu}(\mu)^2 - I_{\nu+1}(\mu)I_{\nu-1}(\mu) ~,
\eeq
which gives 
\beq
\Sigma_{\nu}(\mu) ~=~ \frac{1}{2}\left\{\frac{I_{\nu}(\mu)
I_{\nu+1}(\mu) - I_{\nu-1}(\mu)I_{\nu+2}(\mu)}{I_{\nu}(\mu)^2 - I_{\nu+1}(\mu)
I_{\nu-1}(\mu)} + \frac{\nu}{\mu}\right\} ~.\label{sigmanu2}
\eeq
We note again the presence of an explicit $\nu/\mu$-term from the zero
modes for $\nu\geq 0$ (in which case the first term in ({\ref{sigmanu2})
indeed remains finite in the limit $\mu\to 0$). This term will always be 
present in $\Sigma_{\nu}(\mu)$, for any $N_f$.
It reads in general $\nu/(N_f\mu)$, and thus contributes
$$
\frac{1}{N_{f}}\frac{1}{\mu}\sum_{\nu=-\infty}^{\infty}\nu Z_{\nu}
(\theta=0;\mu) 
$$
to $\Sigma(\theta=0;\mu)$. Using the symmetry (\ref{Zsym}), this is
seen to vanish: At $\theta=0$ this contribution to 
$\Sigma(\theta;\mu)$ always sums up to zero, for any $N_f$. When
$\theta\neq 0$ this is no longer true:
\beqn
\frac{1}{N_{f}}\frac{1}{\mu}\sum_{\nu=-\infty}^{\infty}\nu e^{i\nu\theta}
Z_{\nu}(\theta;\mu) &=& \frac{-i}{N_{f}}\frac{1}{\mu}
\frac{\partial}{\partial\theta} \sum_{\nu=-\infty}^{\infty} e^{i\nu\theta}
Z_{\nu}(\mu) \cr
&=& \frac{-i}{N_{f}}\frac{1}{\mu}
\frac{\partial}{\partial\theta}Z(\theta;\mu) ~.
\label{sigmazeromode}
\eeqn
In particular, for the case at hand, $N_f=2$, we find that this equals
$$
\frac{i\sin(\theta)}{2\mu(1+\cos(\theta))}I_2(\mu(2(1+\cos(\theta))^{1/2})
$$
(which indeed vanishes at $\theta=0$). Moreover, we notice that this
term (\ref{sigmazeromode}) is always strictly imaginary.
Because the full sum is guaranteed to be {\em real}, this means that in
some sense even for $\theta\neq 0$ the contribution of this term to
$\Sigma_{\nu}(\theta;\mu)$ is only apparent; it only serves as to remove a 
corresponding imaginary part from the rest of the sum, leaving a remaining 
real answer.

\centerline{\epsfbox{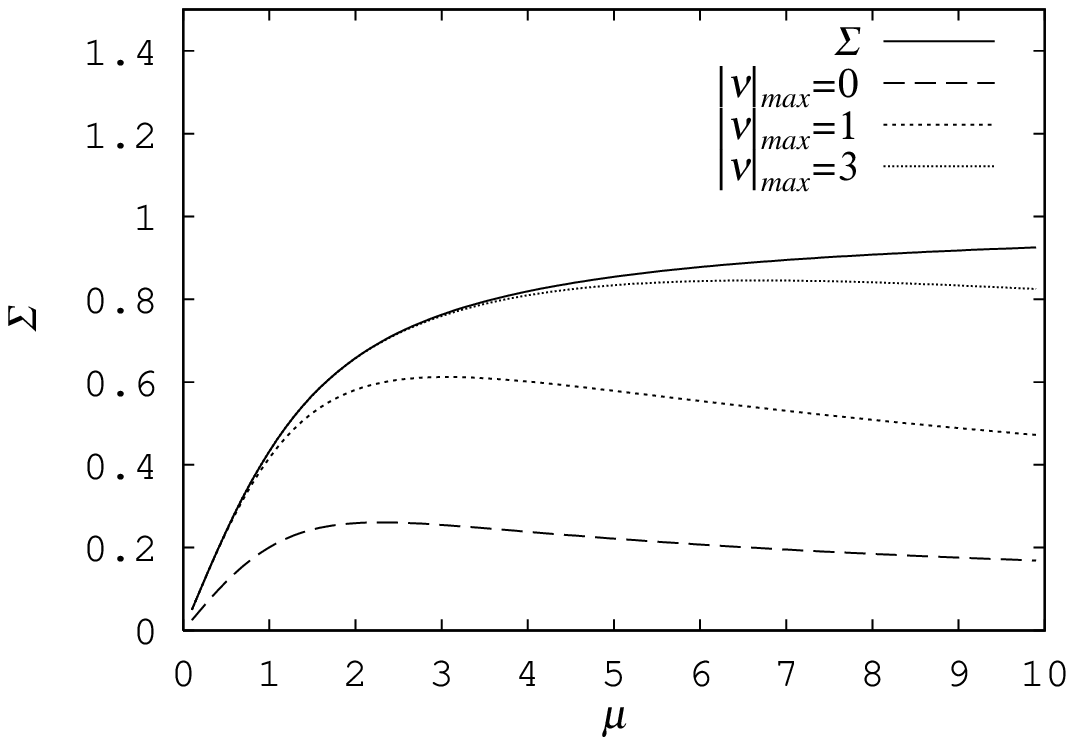}}
{{\small Figure 4: The full chiral condensate 
$\Sigma(\mu)$ for $N_f=2$, $\theta=0$,  and equal masses $\mu$. 
We again compare different approximations, by summing only over the
first few topoloigcal sectors. The $\nu=0$ sector is again not 
particularly significant.}}

\noi
~

\noi
We now check to which degree of accuracy the chiral condensate 
$\Sigma(\theta;\mu)$ for $N_f=2$ and $\theta=0$ ($i.e.$ eq. (\ref{sigma2}))
coincides with just the
chiral condensate in fixed-$\nu$ sectors (\ref{sigmanu2}). This is shown 
in fig. 3 for a few small values of $\nu$. The divergent zero-mode
contributions of course make the condensates for $\nu\neq 0$ very
different from the complete answer. And actually in this $N_f=2$ case the
chiral condensate in just the $\nu=0$ sector at least has the qualitatively
right behavior. But again, to see to what extent the different topological
sectors contribute to the full answer, we must add up the different sectors
with the right weight. In fig. 4 we show how the full chiral condensate
(\ref{sigma2}) is built up as we sum over topological charge $\nu$ from
given $-|\nu|_{max}$ to $+|\nu|_{max}$ in eq. (\ref{sigmanfdef}). 
As in the $N_f=1$ case, we find that the $\nu=0$ contribution
alone is not particularly significant at all. It is obviously a
generic feature, which is not specific to the very special circumstances of 
$N_f=1$. The cases with $N_f\geq 3$ are indeed not qualitatively different
\cite{D2}, and we do not display any of them here.

\section{The Microscopic Spectral Density}

\noi
Eigenvalues $\lambda_n$ of the Dirac operator are defined by $i\slash{D}
\phi_n = \lambda_n\phi_n$, and the spectral density is then
\beq
\bar{\rho}(\lambda;\{\mu_i\}) ~\equiv~ 
\langle\sum_n \delta(\lambda-\lambda_n)\rangle ~,
\label{rhodef}
\eeq
where the average includes the sum over topological sectors. Because the 
weighting is done with the fermion determinant, the density depends on
the masses $\mu_i$ too. The double-microscopic limit is taken by keeping
all $\mu_i=V\Sigma m_i$ and $\zeta=V\Sigma\lambda$ fixed as $V\to\infty$.
The definition (\ref{rhodef}) of course extends to averages in fixed-$\nu$ 
sectors as well:
\beq
\bar{\rho}^{(\nu)}(\lambda;\{\mu_i\}) ~\equiv~ \langle\sum_n 
\delta(\lambda-\lambda_n)\rangle_{\nu} \label{rhonudef}
\eeq
We now go to the double-microscopic limit.
As for any other gauge-field expectation value in fixed-$\nu$ sectors, also 
the microscopic spectral density $\bar{\rho}_S^{(\nu)}(\zeta;\{\mu_i\})$ 
can be summed to give the microscopic spectral density of the full theory:
\beq
\bar{\rho}_S(\zeta;\theta,\{\mu_i\}) ~=~ {\cal Z}(\theta;\{\mu_i\})^{-1}
\sum_{\nu=-\infty}^{\infty}e^{i\nu\theta}{\cal Z}_{\nu}(\{\mu_i\})
\bar{\rho}_S^{(\nu)}(\zeta;\{\mu_i\}) ~.\label{rhofull}
\eeq
It is useful to separate out the zero-modes explicitly:
\beqn
\bar{\rho}_S(\zeta;\theta,\{\mu_i\}) &=& {\cal Z}(\theta;\{\mu_i\})^{-1}
\sum_{\nu=-\infty}^{\infty}e^{i\nu\theta}{\cal Z}_{\nu}(\{\mu_i\})\left[
\rho_S^{(\nu)}(\zeta;\{\mu_i\}) + |\nu|\delta(\zeta)\right] \cr
&\equiv& \rho_S(\zeta;\theta,\{\mu_i\}) ~+~ 
\langle\langle |\nu| \rangle\rangle \delta(\zeta) ~,\label{split}
\eeqn
with an obvious definition of $\rho_S(\zeta;\theta,\{\mu_i\})$.

\noi
It has been shown \cite{D0,OTV} that the microscopic spectral density
(\ref{rhonudef}) without zero-mode contributions can be expressed directly 
in terms of the effective partition function,
\beq
\rho_S^{(\nu)}(\zeta;\{\mu_i\}) ~=~ \frac{1}{2}(-1)^{\nu}|\zeta|
\prod_{f=1}^{N_{f}}
(\zeta^2+\mu_f^2)\frac{{\cal Z}_{\nu}^{(N_{f}+2)}(\{\mu_i\},i\zeta,i\zeta)}{
{\cal Z}_{\nu}^{(N_{f})}(\{\mu_i\})} ~, \label{rhopart}
\eeq
where we for clarity have explicitly indicated the number of fermion
species that are involved on the right hand side.
Actually, from the connection to Random Matrix Theory it is also known that
{\em all} microscopic spectral correlators 
$\rho_S^{(\nu)}(\zeta_1,\ldots,\zeta_n;\{\mu_i\})$ can be derived from
just one ``master formula'' \cite{D0},
\beq
K_S^{(\nu)}(\zeta,\zeta';\mu_1,\ldots,\mu_{N_{f}}) ~=~ \frac{1}{2}(-1)^{\nu}
\sqrt{\zeta\zeta'}\prod_{f=1}^{N_{f}}
\sqrt{(\zeta^2+\mu_f^2)(\zeta'^2+\mu_f^2)}~\frac{
{\cal Z}_{\nu}^{(N_{f}+2)}(\{\mu_i\},i\zeta,i\zeta')}{
{\cal Z}_{\nu}^{(N_{f})}(\{\mu_i\})} ~.\label{mf}
\eeq
All spectral correlation functions follow from this single function:
\beq
\rho_S^{(\nu)}(\zeta_1,\ldots,\zeta_n;\{\mu_i\}) ~=~
\det_{a,b} K_S^{(\nu)}(\zeta_a,\zeta_b;\{\mu_i\}) ~,
\label{correlchUE}
\eeq 
where the determinant on the right hand side is taken in terms of the
$n\times n$ matrix in the arguments $\zeta_a,\zeta_b$.

\noi
The full microscopic spectral density without zero-mode contributions,
$\rho_S(\zeta;\theta,\{\mu_i\})$, takes a particularly simple form in
the massless limit. Using ${\cal Z}_{0}(0) = 1$ and  ${\cal Z}_{\nu}(0) = 0$
for all $\nu \neq 0$, we immediately get
\beq
\rho_S(\zeta;\theta,\{\mu_i=0\}) ~=~ \rho_S^{(\nu=0)}(\zeta;\{\mu_i=0\}) ~.
\label{rhofullmassl}
\eeq
That is,
in the massless limit the full microscopic spectral density without zero
modes simply coincides with the corresponding density of the $\nu=0$ sector. 
Note in particular that in this massless limit the full microscopic
spectral density is independent of the vacuum angle $\theta$. This is as
expected: in QCD with at least one massless flavor any $\theta$-angle
can be removed by a chiral rotation.

\noi
The result (\ref{rhofullmassl}) may seem to contradict the fact that in
the $N_f=1$ theory the chiral condensate is $\Sigma(\theta;\mu\!=\!0) 
= \cos(\theta)$ even though the massless limit is taken at finite volume.
Indeed, even at finite volume there is a kind of Banks-Casher relation
for the {\em microscopic} spectral density too,
\beq
\lim_{\mu\to 0}(2\mu)\int_0^{\infty}\!d\zeta 
\frac{\rho_S(\zeta;\theta,\mu)}{\zeta^2+\mu^2}
~=~ \pi\rho_S(0;\theta,0) ~=~ \pi\rho_S^{(\nu\!=\!0)}(0;0) ~=~ 0~.\label{BC}
\eeq
This microscopic spectral density alone can therefore never give rise to a
condensate in the massless finite-volume theory. But the full condensate
potentially also receives contributions from the zero modes, even at zero
mass. This part equals (taking $N_f$ equal masses for simplicity) 
\beqn
(N_f\mu)^{-1}\langle\langle|\nu|\rangle\rangle &=& 
[N_f\mu{\cal Z}(\theta;\{\mu_i\})]^{-1}
\sum_{\nu=-\infty}^{\infty}e^{i\nu\theta}{\cal Z}_{\nu}(\{\mu_i\})|\nu| \cr
&=&  2[N_f\mu{\cal Z}(\theta;\{\mu_i\})]^{-1}
\sum_{\nu=1}^{\infty}\cos(\nu\theta)
{\cal Z}_{\nu}(\{\mu_i\})\nu ~.\label{absnu}
\eeqn
In the limit $\mu\to 0$ this expression vanishes for all $N_f\geq 2$. 
In view of (\ref{BC}) this 
just means that for $N_f\geq 2$ there are no
chiral condensates in the massless limit, in agreement with the fact
that there cannot be spontaneous symmetry breaking in these finite-volume
theories. For $N_f=1$ and $\theta=0$ the sum in (\ref{absnu}) can be 
performed analytically, giving \cite{LS} 
\beq
\mu^{-1}\langle\langle|\nu|\rangle\rangle =
e^{-\mu}[I_0(\mu)+I_1(\mu)] ~,
\eeq
whose limit at $\mu=0$ indeed equals unity, as required in order to obtain
$\Sigma=1$ in that case. In general, for $\theta\neq 0$ we note again that
because of the behavior $I_n(x) \sim x^{|n|}$ as $x\to 0$, only the
$\nu=\pm 1$ terms contribute to the sum when $\mu\to 0$, yielding 
precisely the required $\mu^{-1}\langle\langle|\nu|\rangle\rangle \to
\cos(\theta)$ in the limit $\mu\to 0$.

\noi
Inserting the formula (\ref{rhopart}) into the definition of 
 $\rho_S(\zeta;\theta,\{\mu_i\})$ in eq. (\ref{split}) gives an interesting
compact expression for the full microscopic spectral density:
\beqn
\rho_S(\zeta;\theta,\{\mu_i\}) &=& {\cal Z}^{(N_{f})}(\theta;\{\mu_i\})^{-1}
\sum_{\nu=-\infty}^{\infty}e^{i\nu\theta}{\cal Z}_{\nu}(\{\mu_i\})
\rho_S^{(\nu)}(\zeta;\{\mu_i\}) \cr
&=& \frac{1}{2}|\zeta|\prod_{f=1}^{N_{f}}(\zeta^2+\mu_f^2)
{\cal Z}^{(N_{f})}(\theta;\{\mu_i\})^{-1}
\sum_{\nu=-\infty}^{\infty}e^{i\nu\theta}(-1)^{\nu}{\cal Z}_{\nu}^{(N_{f}+2)}
(\{\mu_i\},i\zeta,i\zeta) \cr
&=& \frac{1}{2}|\zeta|
\prod_{f=1}^{N_{f}}
(\zeta^2+\mu_f^2)\frac{{\cal Z}^{(N_{f}+2)}(\theta + \pi;\{\mu_i\},
i\zeta,i\zeta)}{{\cal Z}^{(N_{f})}(\theta;\{\mu_i\})} ~.\label{fullrho}
\eeqn
While this formula strongly resembles the one for fixed $\nu$, 
we note the effect of the $(-1)^{\nu}$-factor in the individual
$\rho_S^{(\nu)}(\lambda;\{\mu_i\})$-contributions: The partition function 
with two additional fermion species (in the numerator) is to be evaluated
at a shifted vacuum angle: $\theta+\pi$. Thus even if we insist on getting
the physical microscopic spectral density with $\theta=0$ (or at least
exceedingly small), we need to know also the effective QCD partition function
with two more flavors, and with a non-vanishing vacuum angle.    

\noi
One can perform a similar summation over topological charge
of all microscopic spectral correlators 
$\rho_S^{(\nu)}(\zeta_1,\ldots,\zeta_n;\{\mu_i\})$. However, the 
representation (\ref{mf}) does not give a a convenient compact expression,
due to the fact that the determinant must be taken inside the summation
over $\nu$. Fortunately, there exists a different expression for 
$\rho_S^{(\nu)}(\zeta_1,\ldots,\zeta_n;\{\mu_i\})$ (see the third paper
of ref. \cite{D0}), which, although more unwieldy in general when $\nu$
is fixed, is more convenient in this context:
\beqn
\rho_S^{(\nu)}(\zeta_1,\ldots,\zeta_k;\{\mu_i\}) 
&=& C^{(k)} 
\prod_i^k\left( \zeta_i
\prod_{f=1}^{N_f}(\zeta_i^2+\mu_f^2)\right)
\prod_{j<l}^k(\zeta_j^2-\zeta_l^2)^2 \nonumber\\
&&\times\
\frac{{\cal Z}_{\nu}^{(N_{f}+2k)}
(\{\mu_i\},\{i\zeta_1\},\ldots, \{i\zeta_k\})}
{{\cal Z}_{\nu}^{(N_{f})}(\{\mu_i\})} ,
\label{corrft}
\eeqn
Summing this expression over $\nu$ will yield the microscopic spectral
correlators of the full theory, now expressed in terms of the ratio of a
partition function of $N_f+2k$ species and the usual partition function
corresponding to the actual number of fermions $N_f$. 

\centerline{\epsfbox{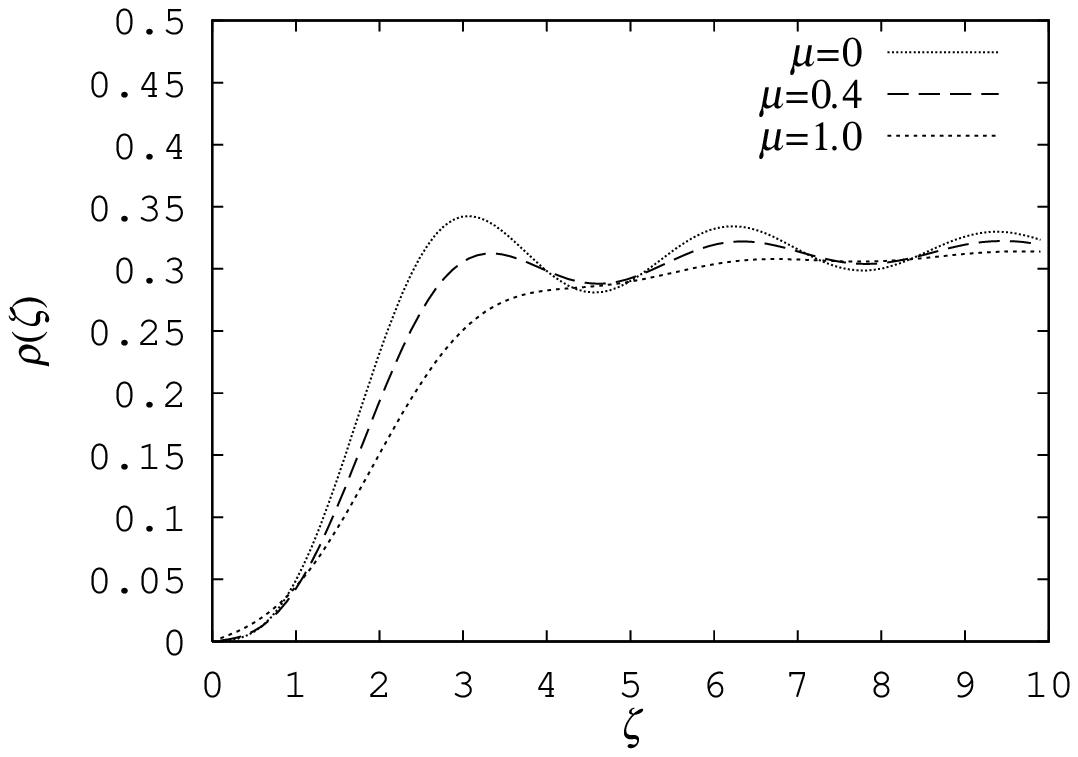}}
{{\small Figure 5: The full microscopic spectral density of the $N_f=1$
theory with $\theta=0$. The massless case coincides with the $\nu=0$
result.}}

\noi
~

\noi
The full effective partition function (\ref{Zeff}) is unfortunately not known
in closed analytical form for $N_f\geq 3$. So to evaluate the right hand
side of eq. (\ref{fullrho}) it is most convenient to simply
perform the sum over topological charges numerically. 
We know already the full answer
in the massless case (\ref{rhofullmassl}), and it is thus interesting to
trace the change in $\rho_S(\lambda;\theta,\{\mu_i\})$ as the mass is
increased. In fig. 5 we show the full microscopic spectral density for
the $N_f=1$ theory with $\theta=0$. The $\mu=0$ curve indeed coincides
exactly with the massless prediction of just the $\nu=0$ sector \cite{V}
\beq
\rho_S^{(\nu=0)}(\zeta;\mu\!=\!0) ~=~ \frac{1}{2}|\zeta|\left[J_{1}(\zeta)^2
- J_{2}(\zeta)J_{0}(\zeta)\right] ~.\label{nf1limit}
\eeq
As the mass increases, sectors with higher values of $|\nu|$ begin to
contribute. Because the oscillations in $\rho_S^{(\nu)}(\zeta;\mu)$ are
shifted to the right as $|\nu|$ increases, there is a certain amount of
destructive interference, and the full microscopic spectral density becomes
smoother as $\mu$ is increased. The limit $\mu \to \infty$, which naively
should correspond to the quenched theory, is ill-defined, as will be 
discussed below.

\centerline{\epsfbox{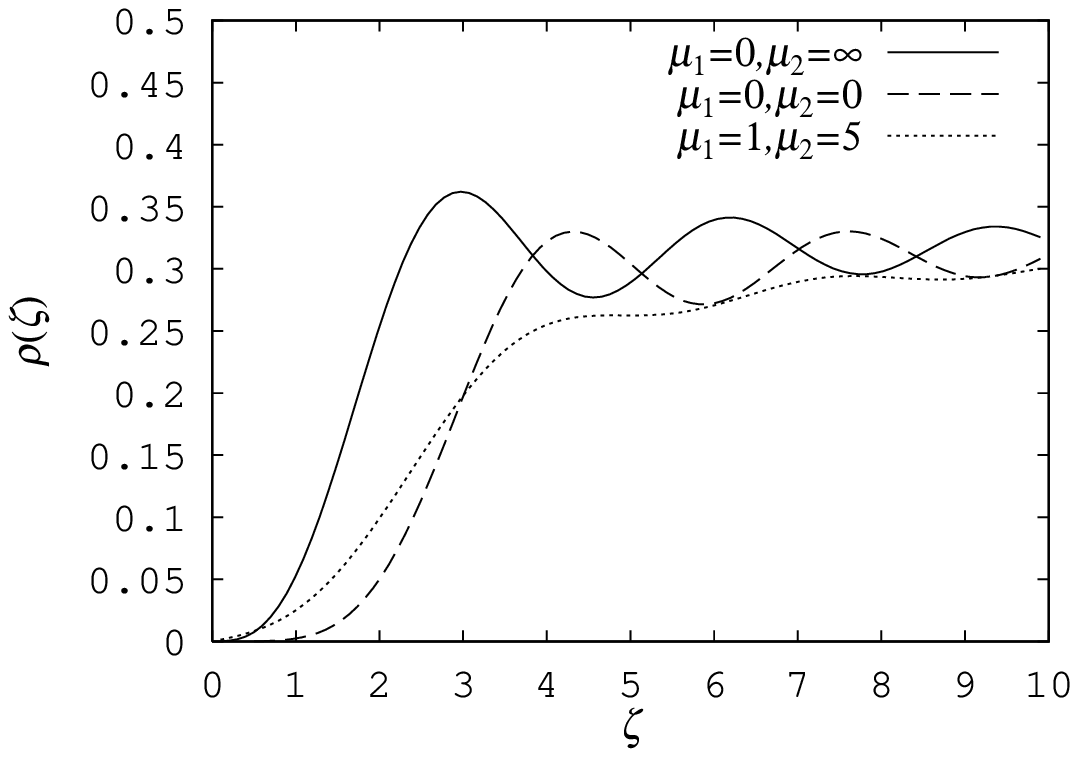}}
{{\small Figure 6: The full microscopic spectral density of the $N_f=2$
theory with $\theta=0$. When both fermion masses are set to zero we
recover the $\nu=0$ result. When one of the masses is taken to infinity, we
recover the full microscopic spectral density of the $N_f=1$, by decoupling.}}

\noi
~

\noi
The case $N_f=2$ is more interesting. Because of the simple identification
(\ref{rhofullmassl}), the $\mu_1=\mu_2=0$ limit simply reproduces the
massless $N_f=2$ density of the just the $\nu=0$ sector,
\beq
\rho_S^{(\nu=0)}(\zeta;\mu_1\!=\!\mu_2\!=\!0) ~=~ 
\frac{1}{2}|\zeta|\left[J_{2}(\zeta)^2
- J_{3}(\zeta)J_{1}(\zeta)\right] ~.
\eeq
In another extreme limit, when, say, $\mu_2\to\infty$, this second fermion
{\em decouples}. If we still take $\mu_1=0$, we thus recover the simple 
massless prediction of just the $\nu=0$ sector of the $N_f=1$ theory, eq. 
(\ref{nf1limit}). For intermediate masses, there is again some amount of
destructive interference that smoothes out the microscopic spectral
density. Shown in fig. 6 are the two limits mentioned above, and one
example of the prediction for intermediate masses.

\subsection{\sc Careful with the sign of $\nu$}

\noi
We have already noticed that the symmetry (\ref{Zsym}) under $\nu\to-\nu$
also implies $\Sigma_{\nu}(\{\mu_i\})=\Sigma_{-\nu}(\{\mu_i\})$. But this 
holds only
if we consider {\em all} contributions, including the $\nu/\mu$-piece.
For example, for $N_f=1$ (and $\theta=0$) the two terms on the right
hand side of eq. (\ref{sigmanu1}) are not separately invariant under
$\nu\to-\nu$, but the sum is. This has consequences for our definition
of the microscopic spectral density $\rho_S^{(\nu)}(\lambda,\{\mu_i\})$,
which does {\em not} include the contribution from zero modes.

\noi
It follows  from eq. (\ref{Zsym}) and the representation (\ref{rhopart})
that also
\beq
\rho_S^{(\nu)}(\lambda,\{\mu_i\}) ~=~ \rho_s^{(-\nu)}(\lambda,\{\mu_i\}) ~.
\eeq
Clearly also the full microscopic spectral density (\ref{rhonudef}) 
is thus symmetric
under $\nu\to-\nu$. Let us for illustration consider one of the ``massive
spectral sum rules'' for $N_f=1$, which was derived in ref. \cite{D2} for
the case $\nu\geq 0$. First, by definition,
\beqn  
\Sigma_{\nu}(\mu) &=& 2\mu\int_0^{\infty}\!d\lambda \frac{
\bar{\rho}_S^{(\nu)}(\lambda;\mu)}{\lambda^2+\mu^2}\cr
&=& 2\mu\int_0^{\infty}\!d\lambda \frac{
\rho_S^{(\nu)}(\lambda;\mu)}{\lambda^2+\mu^2} + \frac{|\nu|}{\mu} ~.
\eeqn
Comparing this with eq. (\ref{sigmanu1}), we see that there are actually two
different massive sum rules, depending on whether $\nu$ is positive or
negative:
\beqn
&&\nu\geq 0:~~~~~~~~\int_0^{\infty}\!d\lambda \frac{
\rho_S^{(\nu)}(\lambda;\mu)}{\lambda^2+\mu^2} ~=~ \frac{I_{\nu+1}(\mu)}{
2\mu I_{\nu}(\mu)} ~,\cr
&&\nu\leq 0:~~~~~~~~\int_0^{\infty}\!d\lambda \frac{
\rho_S^{(\nu)}(\lambda;\mu)}{\lambda^2+\mu^2} ~=~ \frac{I_{\nu+1}(\mu)}{
2\mu I_{\nu}(\mu)} + \frac{\nu}{\mu^2} ~.\label{sumrule1}
\eeqn
It is straightforward to check that in the limit $\mu\to 0$ these two 
relations precisely lead
to two massless spectral sum rules that can be combined into (the sum
running over strictly positive eigenvalues only): 
\beq
\left\langle \sum_n~^{\!\!'}\frac{1}{\lambda_n^2}\right\rangle_{\nu}
~=~ \frac{1}{4(|\nu|+1)} ~,
\eeq
in complete agreement with ref. \cite{LS}. These considerations of course 
trivially generalize to any $N_f$, where only the right hand side of eq. 
(\ref{sumrule1}) becomes increasingly more complicated \cite{D2}.

\noi
The $N_f=1$ case that we have used as an example here is particularly
simple also when it comes to an analytical check on our formula for the
full microscopic spectral density (\ref{fullrho}). In that case one can
explicitly confirm the relation
\beq
\Sigma(\theta;\mu) ~=~ 2\mu\int_0^{\infty}\!d\zeta
\frac{\rho_S(\theta;\zeta,\mu)}{\zeta^2+\mu^2} + \frac{1}{\mu}\langle\langle
|\nu|\rangle\rangle
\eeq
since all the involved integrations can be performed analytically. (The
left hand side is of course just the trivial constant $\cos(\theta)$ in this 
case, and the full spectral density $\rho_S(\theta;\zeta,\mu)$ is indeed just 
a function that achieves this $\mu$-independence of the right hand side as
well).

\subsection{\sc Trouble with the quenched limit}

The quenched limit of $N_f \to 0$ is not easily taken. In gauge-field
sectors of fixed topological charge $\nu$, the quenched mass-dependent
chiral condensate reads \cite{Jac1,OTV}
\beq
\Sigma_{\nu}(\mu) ~=~ \mu\left(I_{\nu}(\mu)K_{\nu}(\mu) + I_{\nu+1}(\mu)
K_{\nu-1}(\mu)\right) + \frac{\nu}{\mu} ~, \label{quecond}
\eeq
where $K_n(x)$ is the modified Bessel function. This quenched chiral
condensate also satisfies $\Sigma_{\nu}(\mu) = \Sigma_{-\nu}(\mu)$. It 
shares the $\sim 1/\mu$ divergence with the 
unquenched condensates when $\nu\neq 0$. In the latter case we have already
seen that this divergence is harmless in the sense that it is killed once
we sum over topological charge. Here this is no longer true. If we 
simply assume
that ${\cal Z}_{\nu}(\mu) = 1$ for all $\nu$ in the quenched limit, it is 
not even possible to sum over topological charge. There is now no suppression 
of higher-$\nu$ contributions, and the sum in
\beq
\Sigma(\theta;\mu) ~=~ \frac{1}{{\cal Z}(\theta;\mu)}
\sum_{\nu=-\infty}^{\infty}e^{i\nu\theta}\Sigma_{\nu}(\mu) \label{qsum}
\eeq
does not converge. If one tries to ``regularize'' the sum by introducing
some maximum topological charge $|\nu|_{max}$ and defining
\beq
{\cal Z}(\theta;\mu)_{reg} ~\equiv~ \sum_{\nu=-|\nu|_{\max}}^{
\nu=+|\nu|_{\max}}e^{i\nu\theta}\Sigma_{\nu}(\mu) ~,
\eeq
eq. (\ref{qsum}) still has no well-defined limit as $|\nu|_{max}\to\infty$. 
This failure of convergence
is not related to the $\mu\to 0$ singularities in fixed-$\nu$ sectors; it
arises here from the infinite summation over topological charge, if, in
the quenched approximation, we assume that each topological sector enters 
with equal weight ${\cal Z}_{\nu}(\mu) = 1$. It is clearly invalid to make 
this assumption. In fact, in the pure gauge theory the topological
susceptibility $\chi_t = \langle\langle \nu^2\rangle\rangle$ (in our units) is 
a well-defined and finite number, and the distribution of topological charge
is believed to be Gaussian in the infinite-volume limit \cite{Luescher}:
\beq
\frac{{\cal Z}_{\nu}}{{\cal Z}} ~\sim~ 
\frac{1}{2\pi\chi_t}e^{-\nu^2/2\chi_{t}} ~.\label{quenchedchi}
\eeq
It is tempting to simply insert this {\em ansatz} for ${\cal Z}_{\nu}$
into eq. (\ref{qsum}) instead of unity, but it is not quite consistent. 
The whole framework here is based on chiral Ward identities, of which \cite{C}
\beq
N_f \langle\langle\nu^2\rangle\rangle ~=~ \mu
\eeq
indeed is a direct consequence of these finite-volume partition functions
\cite{LS}. Taking the quenched limit detaches us from this tight constraint,
and a different analysis is required to establish the behavior of
${\cal Z}_{\nu}$ as a function of $\nu$. In any case, the $\mu\to 0$ 
singularities are no longer removed by the sum over topological charge,
because the distribution of ${\cal Z}_{\nu}$ will be $\mu$-independent. 
This $\sim 1/\mu$ singularity in the (full) quenched chiral condensate has
recently been confirmed by direct lattice measurements using both domain-wall
and overlap fermions \cite{Chen}.

\noi
With these observations, we should certainly expect difficulties with the
quenched microscopic spectral density. The quenched microscopic density in 
a gauge field sector of fixed $\nu$ reads \cite{V,Jac1}
\beq
\rho_S^{(\nu)}(\zeta) = \frac{1}{2}|\zeta|\left[J_{\nu}(\zeta)^2
- J_{\nu+1}(\zeta)J_{\nu-1}(\zeta)\right] ~.
\eeq
According to eq. (\ref{split}), and if we again take ${\cal Z}_{\nu}(\mu)$
to be a $\nu$-independent constant (say, unity), the sum in
\beq
\rho_S(\lambda;\theta,\{\mu_i\}) ~=~ {\cal Z}(\theta;\{\mu_i\})^{-1}
\sum_{\nu=-\infty}^{\infty}e^{i\nu\theta}{\cal Z}_{\nu}(\{\mu_i\})
\rho_S^{(\nu)}(\lambda;\{\mu_i\}) 
\eeq
again requires regularization. Introducing a maximum topological charge
$|\nu|_{max}\to\infty$ as above, we actually find a remarkable regularity. 
Let us for convenience restrict ourselves to $\theta=0$. Shown in fig. 7
is a series of successive summations with increasing $|\nu|_{max}$. The
resulting function is approximately {\em linear}\footnote{I thank S. Nishigaki
for first pointing this out to me.}, until, at a point that
shifts to infinity with increasing $|\nu|_{max}$, it turns over into
the constant value $1/\pi$. The slope of the linear part thus depends on the 
choice of $|\nu|_{max}$, and there is no convergence of the sum. As 
discussed above, the approximation of taking all ${\cal Z}_{\nu}$ to unity 
is just drastically wrong.

\centerline{\epsfbox{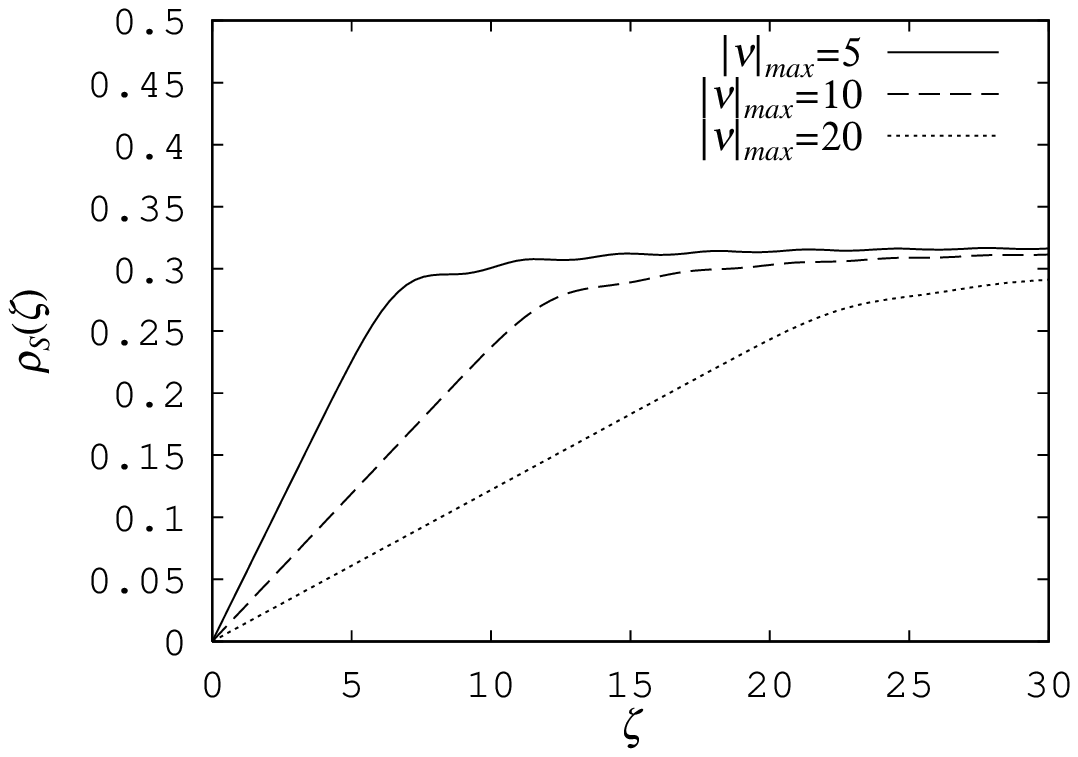}}
{{\small Figure 7: Summing over topological sectors with incorrect weight:
if in the quenched theory one simply assumes that ${\cal Z}_{\nu}=1$
the resulting microscopic spectral density becomes linear up to the
value $1/\pi$, where it turns flat. This point where the linear part
is replaced by a flat part moves to infinity with increasing $|\nu|_{max}$,
and thus has no limit.}}

\noi
~

\noi
Another way of reaching the quenched limit is by decoupling all $N_f$ fermions
in any given theory by sending all masses to infinity. For simplicity taking
$N_f=1$, and sending the mass $\mu$ to infinity, we obtain precisely the
same kind of curve as obtained above in the naive way of just summing
over each fixed-$\nu$ sector with equal weight. The cut-off which controls
the point where $\rho_S(\zeta)$ 
turns from linear to flat behavior is now played
by this mass $\mu$. This suggests that decoupling {\em all} fermions by taking
all their rescaled masses $\mu_i$ to infinity is not providing a physical 
limit of the full theory. Indeed, one easily checks that in this extreme 
limit all partition functions ${\cal Z}_{\nu}$ become independent of $\nu$, 
thus explaining why the two different ways of trying to obtain the quenched 
microscopic spectral density described here fail in precisely the same
manner.  

\noi
How does this result fit in with the general formula (\ref{fullrho})?
Taking the $N_f\to 0$ limit in that equation gives
\beq
\rho_S(\lambda;\theta) 
= \frac{1}{2}|\zeta|\frac{{\cal Z}^{(2)}(\theta + \pi;
i\zeta,i\zeta)}{{\cal Z}^{(0)}(\theta)} ~.
\eeq
For the quenched microscopic density we thus need the effective partition
function for $N_f=2$, which fortunately is known in analytical form (see
eq. (\ref{Z2})). For simplicity taking the case of $\theta=0$, we indeed
find from eq. (\ref{Z2}) that
\beq
\rho_S(\lambda;\theta\!=\!0) ~=~ \frac{1}{2{\cal Z}^{(0)}}|\zeta| ~.
\eeq
The density consistently comes out linear, and again with an undetermined
constant in front. To determine this constant we would normally impose
the matching condition of $\rho_S(\zeta\to\infty) = 1/\pi$, an impossible
requirement in this case. Again, the failure of these attempts at taking
the quenched limit in this framework is perhaps not surprising. The presumed 
distribution (\ref{quenchedchi}) concerns by construction only the pure 
gauge sector, for which no information can be gained via the (anomalous) 
chiral Ward identities on which the present framework is based. Moreover,
in the pure gauge theory we can sum over gauge field configurations with
also fractional winding numbers \cite{'t}. In the large-$N_c$ limit
the topological susceptibility in the quenched theory is, via the
Witten-Veneziano relation \cite{Witten}, related to the flavor singlet
$\eta'$ mass. In the quenched limit, where the anomaly plays no role,
the $\eta'$ should be treated as a (pesudo)-Goldstone boson too. Perhaps
an analogue of the large-$N_c$ effective Lagrangian with built-in
information about the topological susceptibility of the pure gauge theory
(as required by the Witten-Veneziano relation, at least at large $N_c$) can
be used to treat this case in a consistent manner. 

\centerline{\epsfbox{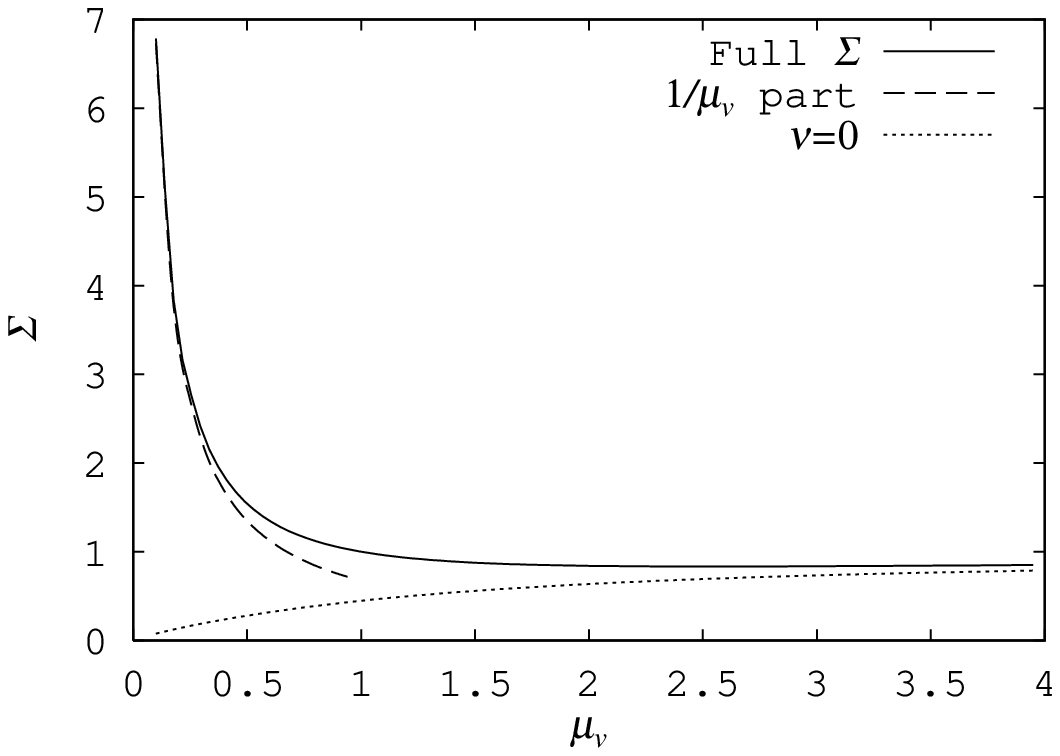}}
{{\small Figure 8: The partially quenched chiral condensate $\Sigma(\mu_v,
\mu)$ in the $N_f=1$ theory with $\theta=0$. The ``quenched'' mass is
indicated by $\mu_v$, while $\mu$ is the real physical mass, here taken
to be unity for convenience. The sum over topological charges converges
everywhere, except at the point $\mu_v=0$. In this case the divergence
at $\mu_v\to0$ can easily be worked out analytically (see eq. 
(\ref{oneovermu})),
and for comparison we display also just the resulting $1/\mu_v$ term with the
coefficient as given in eq. (\ref{oneovermu}).}}

\noi
~

\noi
There are no such difficulties with partial quenching. For example, the
dependence of the partially quenched chiral condensate on the (quenched)
``valence quark'' mass $\mu_v$ is computed in the full theory by averaging
with respect to the usual fixed-$\nu$ partition function ${\cal Z}_{\nu}$.
This sum converges for $\mu_v\neq 0$. 
To be concrete, consider the analytical expression for
the $N_f=1$ case which was derived in the second of ref. \cite{OTV}:
\beqn
\Sigma_{\nu}(\mu_v,\mu) &=& \mu_v\left[I_{\nu+1}(\mu_v)K_{\nu+1}(\mu_v)
+ I_{\nu+2}(\mu_v)K_{\nu}(\mu_v)\right] \cr
&&+ 2\mu\frac{K_{\nu}(\mu_v)}{I_{\nu}(\mu)}\frac{\mu_vI_{\nu}(\mu_v)
I_{\nu+1}(\mu)-\mu I_{\nu}(\mu)I_{\nu+1}(\mu_v)}{\mu_v^2-\mu^2} + \frac{\nu}
{\mu_v}
\eeqn
In fig. 8 we show this as a function of $\mu_v$ for 
$\nu=0$ (which of course does not exhibit the $1/\mu_v$-singularity), and 
fixed value of the physical ``sea quark'' mass $\mu$. We also
show the full answer, for the same mass $\mu$, obtained by summing over
winding numbers according to
\beq
\Sigma(\theta;\mu_v,\mu) ~=~ {\cal Z}(\theta;\mu)^{-1}
\sum_{\nu=-\infty}^{\infty}e^{i\nu\theta}
{\cal Z}_{\nu}(\mu)\Sigma_{\nu}(\mu_v,\mu) ~,
\eeq
where in this case ${\cal Z}(\theta;\mu)=\exp[\mu\cos\theta]$, and
${\cal Z}_{\nu}(\mu)=I_{\nu}(\mu)$.

\noi
The partially quenched cases are particularly interesting because they
provide a limit in which the $\sim 1/\mu_v$ divergences of the quenched
approximations are under full control from the analytical point of view.
Indeed, the coefficient of the $1/\mu_v$-term in the partially quenched 
$N_f=1$ theory can be computed analytically in this finite-volume limit;
it equals
\beq
\langle\langle|\nu|\rangle\rangle ~=~ \mu e^{-\mu}(I_0(\mu)+I_1(\mu)) ~.
\label{oneovermu}
\eeq
This curve has also been included in fig. 8, and one clearly sees how
accurately it describes the $1/\mu_v$ singularity.

\subsection{\sc Extensions}

\noi
The present results easily extend to the other two major classes of chiral 
symmetry breaking \cite{Jac}. The analytically most simple case is that 
corresponding to gauge group SU($N_c\geq 3$) and $N_f$ fermions in the adjoint
representation (which in Random Matrix Theory language corresponds to the
chSE). For example, from the relations derived in the second of reference
\cite{D0} we know that we in this case can express the microscopic spectral 
density of the Dirac operator in a sector of topological charge $\nu$ by
\beq
\rho_S^{(\nu)}(\zeta;\{\mu_i\}) ~=~ C_4~ \zeta^3~
\frac{{\cal Z}^{(N_{f}+4)}_\nu(\{\mu_i\},\{i\zeta\})}
{{\cal Z}^{(N_{f})}_\nu(\{\mu_i\})} ~.
\label{symquenched}
\eeq
Also this can be summed over topological charge $\nu$ to get the full 
microscopic spectral density for that case. 
Using a similar relation for all higher $k$-point functions \cite{D0}
\beqn
\rho_S^{(\nu)}(\zeta_1,\ldots,\zeta_k;\{\mu_i\}) 
&=& C_{4}^{(k)} 
\prod_i^k\left( \zeta_i^{3}
\prod_f^{N_f}(\zeta_i^2+\mu_f^2)\right)
\prod_{j<l}^k|\zeta_j^2-\zeta_l^2|^{4} \nonumber\\
&&\times\
\frac{{\cal Z}_{\nu}^{(N_{f}+4k)}
(\{\mu_i\};\{i\zeta_1\},\ldots, \{i\zeta_k\})}
{{\cal Z}_{\nu}^{(N_{f})}(\{\mu_i\})} ,
\label{corrft4}
\eeqn
also these $k$-point functions can be averaged over topological charges
to yield a ratio of two full partition functions, without restrictions to
topology. In these relation each (imaginary) additional fermion mass on
the right hand side is four-fold degenerate. Unfortunately, the relevant
fixed-$\nu$ partition functions are not in general known in closed analytic
form, and the proportionality constants in both of the above relations
have therefore not yet been determined. Until this has been done, and
in particular their $\nu$-dependences have been found, we cannot 
explicitly identify
the combination of finite-volume partition functions that will emerge from
the sums.

\section{Conclusions}

\noi
We have here considered the effect of summing over all topological charges
$\nu$, while remaining in the finite-volume scaling regime defined by eq.
(\ref{meso}). For the mass-dependent chiral condensate the effect of summing
over topological sectors is quite drastic. For example, the condensate
actually diverges like $\sim 1/\mu$ in the $\mu\to 0$ limit if one restricts
oneself to any sector of non-vanishing topological charge $\nu$, even though
the full condensate vanishes as $\mu\to 0$ (except for the case $N_f=1$).
The microscopic Dirac operator spectrum itself is also substantially
modified by the summation over topological charges. A simple compact relation
gives the full microscopic spectral density in terms of a ratio of
partition functions, one of which is evaluated at a shifted vacuum angle
$\theta\to \theta+\pi$, and with two additional fermion species. We have
pointed out a difficulty with obtaining the quenched ($N_f\to 0$) limit in 
this simple way. The difficulty disappears if one focuses instead
on a partially quenched limit, in which the appropriate weight factors
in the sum over topological numbers again are known exactly in this 
finite-volume regime.
We have also noted a drastic simplification in the massless
case, where the full microscopic spectral density actually coincides with
the one of the sector with $\nu=0$.

\noi
An interesting open question is whether one can establish a compact
Random Matrix Theory formulation also for the full effective partition
function ${\cal Z}(\theta;\{\mu_i\})$. Although we believe the answer
to be negative, we are not aware of any strong argument to that effect.

\noi
In future lattice studies it seems obvious to try to study not just the
microscopic Dirac operator spectrum in sectors of fixed $\nu$, but also the 
associated distribution of configurations labelled by the topological index.
Thanks to very recent developments this is now possible \cite{L,Urs}. This
way one can test simultaneously the probability
distribution of configurations with fixed topological index (which is known
analytically in the regime (\ref{meso}) and given, for $\theta=0$, simply by 
${\cal Z}_{\nu}(\{\mu_i\})/{\cal Z}(\theta\!=\!0;\{\mu_i\})$), and, 
simultaneously, the full microscopic Dirac operator spectrum. Such
numerical simulations with dynamical fermions now seem feasible.

\vspace{1cm}
\noi
{\sc Acknowledgements:}~ Discussions with Urs Heller and Jac Verbaarschot
are gratefully acknowledged. This work has been partially supported by
EU TMR grant no. ERBFMRXCT97-0122.


\end{document}